\documentclass{ws-procs975x65}

\begin{document}

\title{LABORATORY TESTS OF GRAVITATIONAL PHYSICS USING A CRYOGENIC TORSION PENDULUM}

\author{E. C. Berg, M. K. Bantel, W. D. Cross, T. Inoue, and R. D. Newman}

\address{University of California at Irvine,\\
4129 Frederick Reines Hall,\\ 
Irvine, CA 92697-4575, USA\\ 
E-mail: eberg@uci.edu}

\author{J. H. Steffen, M. W. Moore, and P. E. Boynton}

\address{University of Washington at Seattle,\\ 
P.O.Box 351560,\\
Seattle, WA 98195-1560, USA\\
E-mail: jsteffen@u.washington.edu}  


\maketitle

\abstracts{
Progress and plans are reported for a program of gravitational physics experiments using cryogenic torsion pendula undergoing large amplitude torsional oscillation.  The program includes a UC Irvine project to measure the gravitational constant {\it G} and joint UC Irvine--U. Washington projects to test the gravitational inverse square law at a range of about 10 cm and to test the weak equivalence principle.}

\section{Introduction}
The torsion pendulum remains the instrument of choice for a variety of laboratory tests of gravitational phenomena.  Such experiments necessarily require the detection of extremely small torques.  In early applications, this instrument was operated in a mode in which the signal torque was detected as an angular deflection of a pendulum in the presence of a source mass.  This ``deflection method" is susceptible to systematic error produced by even a small tilt of the apparatus.  One way to overcome this drawback is to operate with the pendulum executing torsional oscillations and to extract the signal as the measured shift in oscillation frequency correlated with changes in relative orientation between the pendulum equilibrium position and the direction to the source mass\cite{uwmoriond}. 

This ``frequency method" was commonly used in the ${\rm 20^{th}}$ century for measurements of the gravitational constant {\it G}, but with milliradian oscillation amplitude.  We now operate at much larger amplitude, nearer the maximum in signal-to-noise ratio\cite{uwa2}.  A downside to this measurement technique has been the temperature sensitivity of a torsion fiber's elastic modulus.  Variation in fiber temperature will produce systematic and random errors in the frequency measurement.  An effective remedy is to operate the pendulum in a cryogenic environment, providing both reduced temperature sensitivity and an opportunity for improved temperature control.  This is the path pursued in the experiments described here.

Details of our cryogenic torsion pendulum system, which uses this frequency method, can be found elsewhere\cite{mg7,moriond,london,uzbek}${\rm ^,}$\footnote{References 5,8,9,10,11 are available at: {\sf http://www.physics.uci.edu/gravity/}}.  The cryogenic environment has a number of benefits besides temperature stability, including low thermal noise (${\it \sim\sqrt{\frac{k_BT}{Q}}}$), high fiber {\it Q} ($>$ ${\rm 10^5}$), slower drift of angular equilibrium position, easily achieved high vacuum, and effective magnetic shielding with superconducting materials.  Concerns of bias due to fiber nonlinearity and anelasticity have been addressed and appear not to be significant\cite{kuroda,icifuas}.  Our apparatus is located at the Battelle Gravitational Physics Lab on the Department of Energy--Hanford site in the desert of eastern Washington State.  This underground facility on the slope of a basalt mountain is several kilometers from public access and associated anthropogenic noise, and was developed largely because of its low ground-motion background (power density less than ${\rm 10^{-22}}$ ${\rm m^2/Hz}$ at 10 Hz).

\section{G Measurement Progress}
\begin{figure}[ht]
\centerline{\epsfxsize=4in\epsfbox{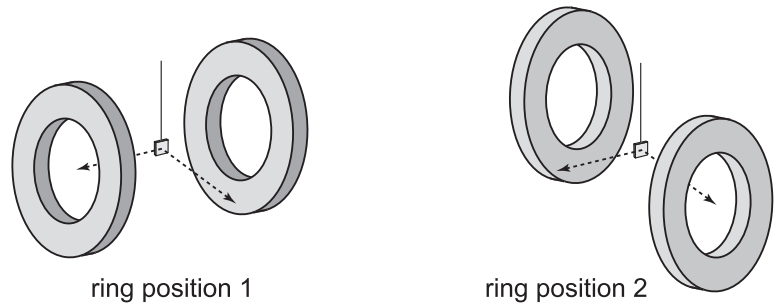}}   
\caption{Copper rings (59 kg each) and fused silica thin plate pendulum (11 g) for the measurement of {\it G}.
\label{rings}}
\end{figure}

\begin{figure}[ht]
\centerline{\epsfxsize=1.5in\epsfbox{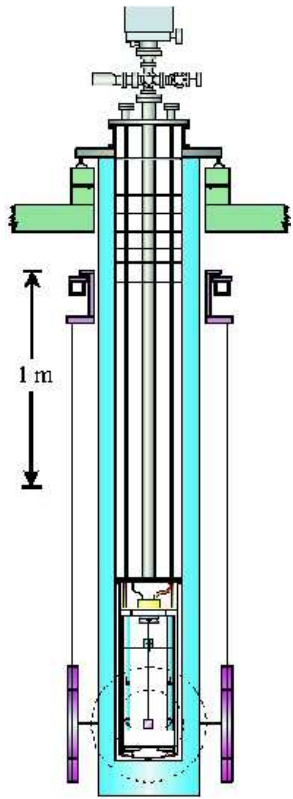}}   
\caption{Experimental cryostat configured to measure {\it G}.
\label{cryostat}}
\end{figure}

The configuration used to measure {\it G} (Figures \ref{rings} and \ref{cryostat}) employs a pair of source mass rings outside the cryogenic environment, which produce an extremely uniform field gradient at the pendulum such that the signal is predominantly due to coupling between the field and the pendulum's quadrupole moment.  The frequency shift due to the source mass rings may be expressed as: ${\it \omega_1^2 - \omega_2^2}$ ${\it =}$ ${\it KG}$ ${\it J_1(2A)/A}$ plus small higher order terms, where ${\it \omega_1}$ and ${\it \omega_2}$ are the pendulum frequencies for the ring positions 1 and 2 indicated in Figure \ref{rings}, {\it K} is a constant determined by geometry and the mass and dimensions of the pendulum and rings, ${\it J_1}$ is a Bessel function, and {\it A} is the oscillation amplitude of the pendulum.  Thus by measuring the frequencies and amplitude of the pendulum, {\it KG} may be determined.  The factor {\it K} is proportional to the ratio of the pendulum's quadrupole moment to moment of inertia, a ratio that for our thin pendulum is highly insensitive to uncertainties in pendulum mass distribution.  The uniformity of the field gradient makes the {\it G} measurement highly insensitive to error in the positioning of the pendulum ({\it e.g.} a 3 mm error produces a ${\it \delta G/G}$ ${\rm <}$ 1 ppm).  We operate the pendulum at an oscillation amplitude near an extremum of {\it $J_1$(2A)/A} (2.57, 4.21, or 7.40 radians) so that the frequency shift is very insensitive to error in amplitude determination.  The higher signal sensitivity at 2.57 radians makes it the natural choice for the majority of our data, and data at the two higher amplitudes is used as a systematic check.  The frequency method requires precise timing measurements of the null signal of an optical lever.  The pendulum period of 135 sec shifts by ${\rm \sim1.5}$ msec due to the source mass modulation, and this shift is measured to ${\rm \sim300}$ nsec for each oscillation cycle. 

\eject

\begin{figure}[ht]
\centerline{\epsfxsize=5in\epsfbox{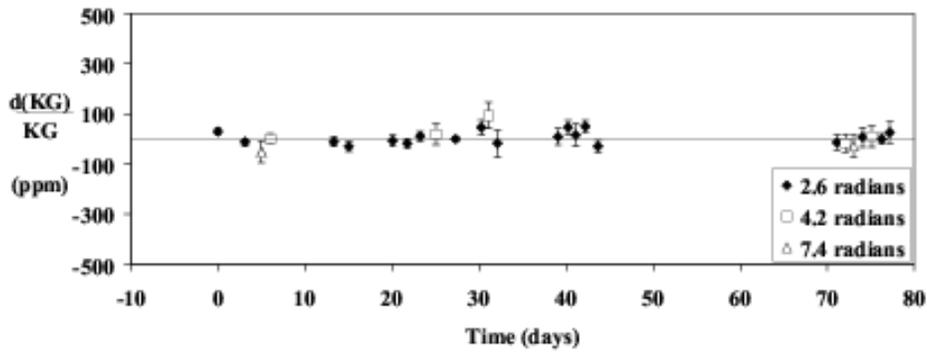}}   
\caption{Year-2000 dataset results for {\it G}, plotted with an arbitrary zero point.  A heat-treated fiber was used after day 50.  Data was taken for three oscillation amplitudes as noted.
\label{results00}}
\end{figure}

\begin{figure}[ht]
\centerline{\epsfxsize=5in\epsfbox{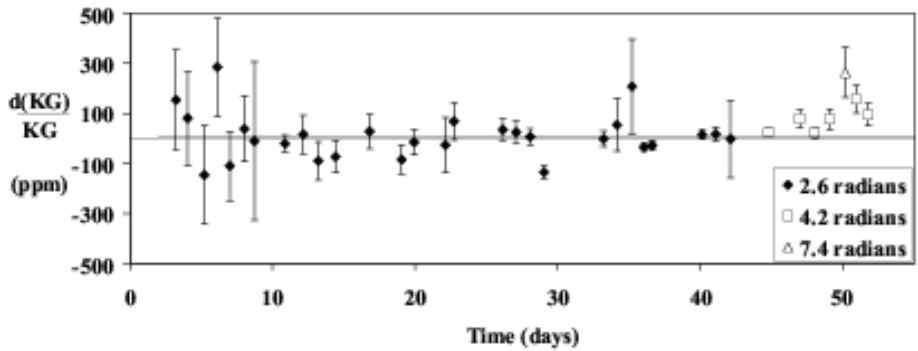}}   
\caption{Year-2002 dataset results for {\it G}, plotted with the same zero point.  The same heat-treated fiber was used as in year 2000.
\label{results02}}
\end{figure}

{\it G} measurements are presented here in terms of {\it KG}, where {\it K} includes an arbitrary and hidden zero point to reduce investigator bias.  Our uncertainty on the value of {\it K} is 7 ppm.  Results from 690 hours of raw data taken in 2000 are reproduced in Figure \ref{results00}\cite{mg9,cpem02}.  Results from 660 hours of raw data taken in 2002 are shown in Figure \ref{results02} on the same scale with the same zero point.  The data taken with 2.57 radian oscillation amplitude yield a weighted average of {\it KG} with 10 ppm (2000) and 18 ppm (2002) statistical uncertainty, consistent within 24 ${\rm \pm}$ 21 ppm. 
 
\eject

In the 2002 dataset, note the increased noise that is believed to be dominantly due to a vacuum leak and poor thermal control due to electronic noise coupled into our control system.  Also note the bias at the end of the dataset, when data was taken with higher oscillation amplitudes.  The origin of this bias, which was not pronounced in the 2000 data, is unclear.  The increased noise problem should be resolved upon completion of various upgrades currently being implemented.  Additional data will be taken mid-2004 using an AL5056 fiber in place of BeCu to test for fiber-related systematic errors. 

\section{Cryogenic Apparatus Upgrades}
\begin{figure}[ht]
\centerline{\epsfxsize=2.5in\epsfbox{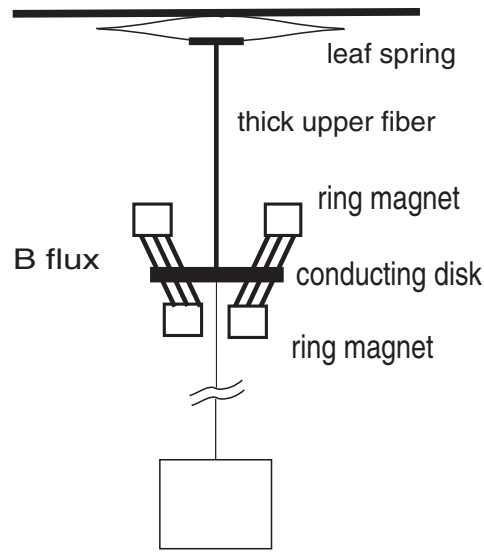}}   
\caption{Improved magnetic damping of bounce and swing modes.  
\label{damping}}
\end{figure}

A redesigned magnetic damping system (Figure \ref{damping}) will minimize not only swinging modes but also vertical stretch ``bounce" modes of the pendulum, and our next dataset should show the benefit of removing this source of noise.  A stiff upper fiber couples to the torsion fiber through an aluminum disk positioned between two ring magnets with an iron return path.  Induced eddy currents in the disk damp swinging motion in two dimensions.  The redesigned system uses ring magnets of different sizes, producing field lines with radial as well as vertical components.  A beryllium-copper leaf spring suspension partially decouples vertical motion of the apparatus from the pendulum, and couples motion of the disk to bounce modes, damping them. 

\eject

Improved temperature control is one of the significant benefits of going to low temperature; typically the pendulum support temperature remains within ${\rm \pm}$100  ${\rm \mu K}$ over a 25 hour run.  Our 2002 dataset however showed ten times worse temperature control.  The cause of the problem is believed to be electrical coupling with our source mass stepper motor and other room temperature electronics.  Recently improved isolation, shielding, and conditioning should eliminate this coupling and restore or improve our previous temperature control for future measurements.

Leaks of helium into the pendulum's vacuum chamber, evident only at low temperature, are suspected of being a major cause for the noise increase observed in the 2002 dataset, and are being repaired.

\begin{figure}[ht]
\centerline{\epsfxsize=2.5in\epsfbox{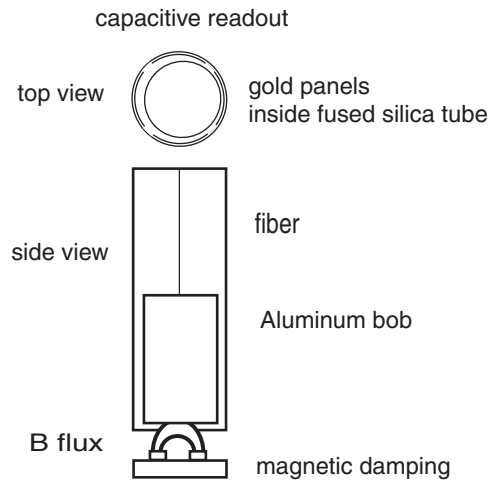}}   
\caption{Capacitive tiltmeter with magnetic swing damping.
\label{tilt}}
\end{figure}

In the case of the balanced source mass configuration for our {\it G} measurement (identical rings hanging on opposite sides), tilt effects are negligible to the degree that our source mass rotation axis is centered between the source masses.  As discussed below, however, our inverse square law and equivalence principle tests rely on a large source mass on one side, and here it is essential to monitor and control tilt.  To this end tiltmeters were developed and installed slightly above and on four sides of the pendulum inside its vacuum chamber (see Figure \ref{tilt}).  The tiltmeters are monitored with a capacitance comparator chip and configured for sensitivity along one axis each.  An active tilt control system will be directed by two of these tiltmeters and be monitored by the other two.  Preliminary tilt measurements at liquid helium temperature indicate an upper limit on the noise level of 500 ${\rm nrad\sqrt{day}}$ at ${\rm \sim0.3}$ mHz.
 
\eject

Heating of the pendulum by the optical lever light source used to determine its angular position produces variations in the fiber temperature and consequently noise in the frequency measurement.  This effect is minimized by activating the light source only for those brief intervals when it gives useful information.  This source of noise is further reduced by adding a higher reflectivity gold coating to the previously aluminum-coated pendulum and by further reducing the ``light-on" duty cycle.  Normally between runs the pendulum is re-excited to a specific amplitude and the motion is reacquired by turning on the optical lever continuously for several periods, with significant resultant heating of the pendulum.  Development of an automated system for re-excitation that maintains a lock on the pendulum's motion will obviate the need for this extra illumination.

Replacement of our 850 nm LED light source with a more powerful 1550 nm incoherent SLED light source is being considered.  This could yield lower shot noise and a net reduction in absorbed light as well.  Split photodiode detectors suitable for this wavelength and a single-mode fiber optic (smaller spot size) are being tested.  Initial tests indicate an unacceptable noise level from the SLED source, however.

A rotary ``piezomotor" is being developed that will use stacks of shear-polarized piezoelements to rotate the pendulum's suspension point relative to the apparatus.  This will enable important tests for systematic errors in our experiments.

Extensive studies have been performed on torsion fibers including tungsten, sapphire, BeCu, heat-treated BeCu, and Al5056\cite{icifuas}.  Our 2000 and 2002 datasets were taken using 20 ${\rm \mu m}$ BeCu fibers (one heat-treated the other not).  A planned {\it G} measurement for Spring 2004 will use a 50 ${\rm \mu m}$ AL5056 fiber to investigate systematic bias due to fiber properties.  Sapphire is attractive because of the high {\it Q} it affords, but is not electrically conductive, posing risk of electrostatic biases.  The thinnest fiber currently available (70 ${\rm \mu m}$) is undesireably stiff.  Attempts to procure thinner fibers or reduce the diameter of a 70 ${\rm \mu m}$ fiber continue.

\section{Equivalence Principle Test Pendulum}
A pendulum for testing the weak equivalence principle is being designed as well, and we hope to begin fabrication soon\cite{ep}.  The experiment will probe for a composition-dependent force at various distance scales; thus, our field sources will be a local source mass, a mountain approximately 1 km away, and the sun.

\begin{figure}[ht]
\centerline{\epsfxsize=2.5in\epsfbox{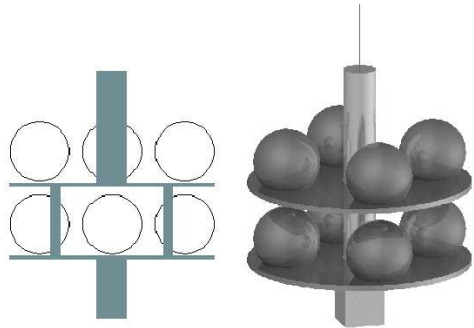}}   
\caption{Current equivalence principle test pendulum design and an earlier rendered design.
\label{pendulum2}}
\end{figure}

In the current design, the pendulum will carry a total of eight spheres approximately 2 cm in diameter.  Four will be beryllium, the other four will be made of a magnesium alloy with density closely matched to Be.  The spheres will all have the same diameter, but in the denser spheres, small holes will be drilled symmetrically at six points around the surface in order to match the masses of the spheres.  Be and Mg were chosen due to their similar densities and to their relatively large difference per unit mass in both binding energy and neutron to proton ratio.  By design, the pendulum will have four-fold mass symmetry, eliminating all Newtonian multipole moments for {\it l} = 1 through {\it l} = 4, except for the (4,4) moment.  There will be three positioning nubs (not shown) for each of the spheres, and trim masses that can slide up and down on the four rods that connect the two support plates.  The spheres will be placed on a frame of fused silica (see Figure \ref{pendulum2}).  On its base will be a four-sided mirror for angle readout with an optical lever.

A Be sample and samples of the Mg alloy that we will use were tested for magnetic contamination by suspending them in a torsion pendulum between poles of a rotating permanent magnet.  The permanent magnetization in all of the samples was found to be less than ${\rm 10^{-9}}$ ${\rm Am^2/cm^3}$.  The measured density of our Mg alloy (1.84 ${\rm g/cm^3}$) differs from that of our Be sample by +0.8\% at ${\rm \sim}$ 2 K. 

\eject

\section{Inverse Square Law Test}
Along with  tests of the equivalence principle it is important to verify the weak-field limit of general relativity in a manner that is composition independent; namely, through tests of the inverse square force law.  An inverse square law violation (``ISLV") could imply the existence of a superposed non-Newtonian force that couples at least approximately to some combination of mass, baryon number, and lepton number. 

We describe a null experiment in which a specially configured torsion pendulum undergoing large-amplitude oscillations in proximity to a source mass also of special form, is potentially able to detect inverse square law violations as small as  ${\rm 10^{-5}}$ of standard gravity at a range around ten centimeters --- a two-order-of-magnitude improvement over the current empirical limit at that length scale.  This improvement is accomplished by an experimental design that is only second-order sensitive to fabrication errors in pendulum and source mass, and not by a substantial reduction in dimensional tolerances. 

\subsection{Experimental Approach}
\begin{figure}[ht]
\centerline{\epsfxsize=3in\epsfbox{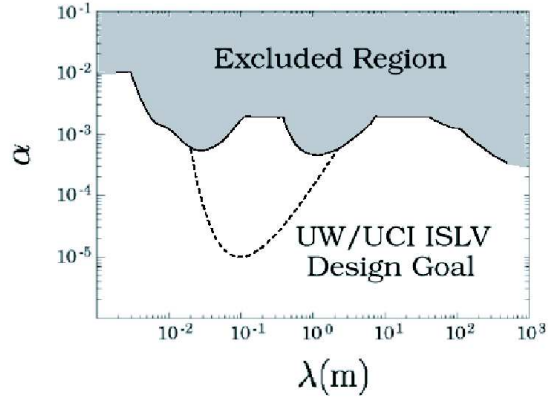}}   
\caption{Composition-independant, experimental, 2${\it \sigma}$ upper bound on ${\it \alpha}$ as a function of ${\it \lambda}$.  The dashed line denotes the design goal of this ISLV experiment.
\label{alphalambda}}
\end{figure}

There are several ways to characterize a violation of the inverse square law.  The most common is to add a Yukawa term to the gravitational potential, giving the interaction potential
\begin{eqnarray}
V&=&V_g+V_Y \nonumber \\[4pt]
{}&=&-\frac{GM}{r}\left(1+\alpha e^{-r/\lambda}\right)\, ,
\label{fullpot}
\end{eqnarray}
where {\it G} (at ${\it r=\infty}$) is Newton's constant, {\it M} the mass of the object generating the potential, ${\it \alpha}$  the strength of the ISLV potential relative to gravity, and ${\it \lambda}$  a characteristic length scale of the interaction.  Figure \ref{alphalambda} shows the current experimental constraint on ${\it \alpha}$ as a function of ${\it \lambda}$.

While a Newtonian gravitational potential satisfies Laplace's equation in a region with no gravitational sources, a non-Newtonian (ISLV) interaction manifests itself as a nonzero Laplacian of the vacuum potential.  However, because the Laplacian is a scalar and thus spherically symmetric, it cannot produce a torque on our pendulum.  We therefore designed our experiment to detect a nonzero horizontal gradient of the Laplacian, the lowest multipole-order indication of ISLV in a torsion experiment\cite{uwpakistan}.  Because we directly detect the horizontal gradient of the Laplacian, our experimental results will be independent of the ISLV model.

The total interaction energy between an object with density profile ${\it \rho(\textbf{{\it r}})}$ and the potential (\ref{fullpot}) is given by
\begin{equation}
U = \int \rho Vd^3r.
\label{energy}
\end{equation}

A Taylor expansion of  the potential in cartesian coordinates about the origin yields
\begin{equation}
U = \int \rho \left( 1 + x\frac{\partial}{\partial x} + y\frac{\partial}{\partial y} + z\frac{\partial}{\partial z} + \frac{1}{2} x^2 \frac{\partial^2}{\partial x^2} + xy \frac{\partial^2}{\partial x \partial y} + \ldots \right)V dxdydz
\label{energyexp}
\end{equation}
where the derivatives are evaluated at the origin.  These terms can be grouped into expressions that have the symmetries of the spherical harmonics.  This gives a concise, orthonormal basis in which to express the interaction energy.  The general expansion of the interaction energy may then be written as
\begin{equation}
U = \sum_{nlm} V_{nlm} M_{nlm}
\end{equation}
where
\begin{equation}
M_{nlm} \propto \int \rho r^n Y_l^m d^3r
\end{equation}
and the ${\it V_{nlm}}$'s are the appropriate derivatives of the potential evaluated at the origin.  Grouping the cartesian terms of Equation (\ref{energyexp}) into the spherical harmonics is nontrivial and normalization of the moments is also challenging.  This problem was examined by Moore\cite{uwa1}.

The horizontal gradient of the Laplacian corresponds to the multipole term
\begin{eqnarray}
U_{311} &=& V_{311}M_{311} \nonumber \\[4pt]
{}&=& \frac{\partial}{\partial x} \left( \nabla^2 V \right) \int \rho x \left(x^2+y^2+z^2 \right) dxdydz \nonumber \\[4pt]
{}& \propto &\frac{\partial}{\partial x} \left( \nabla^2 V \right) \int \rho r^3 Y_1^1 dxdydz\, .
\end{eqnarray}
We call ${\it M_{311}}$ and ${\it V_{311}}$ the 311 mass and field moments respectively.  The 311 mass moment is manifestly non-Newtonian because the field moment to which it couples, ${\it V_{311}}$, is a derivative of the Laplacian and would be identically zero if the interaction were strictly Newtonian.

To take advantage of the multipole expansion, the ratio of the size of the pendulum to the distance to the source mass is less than one, ensuring the contribution to the interaction energy from each higher order is suppressed approximately by that ratio.  Our pendulum and source mass are designed so that the low-order Newtonian mass and field moments are eliminated up to the order for which the contribution from the nonzero moments falls below estimated systematic errors.  In particular, since our signal has {\it m} = 1 symmetry, we null by design all {\it m} = 1 Newtonian pendulum mass moments from {\it l} = 1 to {\it l} = 6 and source-mass field moments from {\it l} = 1 \break to {\it l} = 8.

Historically, experiments designed to detect ISLV using a torsion device have relied on a pendulum with an exaggerated Newtonian moment.  Indeed, a high sensitivity to ISLV could be achieved using a barbell with large {\it l} = 2, {\it m} = 2 mass moment.  On the other hand, this exaggerated Newtonian moment would couple to the residual {\it l} = 2, {\it m} = 2 field moments (arising from  fabrication errors associated with the source mass) to produce a correspondingly large {\it m} = 2 systematic effect that would mimic a putative ISLV signal in such an experiment.  That is, any such scheme to increase the ISLV signal moment would necessarily increase the limiting systematic effect as well.

By designing a pendulum that has no low-order Newtonian moments we can reduce such systematic effects by two orders of magnitude.  Residual Newtonian moments resulting from fabrication errors in the source mass would not directly interact with the pendulum since it is specifically designed not to detect them.  It is therefore only the residual Newtonian mass moments resulting from errors in fabricating the pendulum that can couple to those residual Newtonian field moments.  As a result, departures from the design configuration are manifest only in second order in these errors.  This design strategy dramatically reduces systematic effects arising from standard machining tolerances, metrology, and material density inhomogeneity without resorting to heroic fabrication techniques. 

\subsection{Experimental Design}

\begin{figure}[ht]
\centerline{\epsfxsize=2in\epsfbox{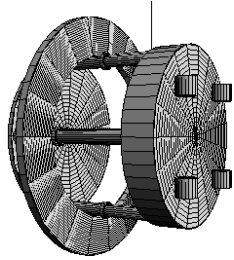}}   
\caption{Rendition of the ISLV pendulum.
\label{islvpend}}
\end{figure}

The result of pursuing this approach to pendulum design is shown in Figure \ref{islvpend}, and this configuration is being fabricated at UC Irvine from fused silica parts.  It has a mass of 240 g and is approximately 8 cm in diameter.  The pendulum is suspended by a BeCu fiber that, under the weight of the pendulum, is stressed to about 60\% of its tensile strength.  The pendulum is gold coated and the fiber  grounded to suppress electrostatic torques (important because the minimum detectable signal energy for {\it S/N} = 1 is only on the order of tens of eV).  The fabrication tolerance of the pendulum parts and their assembly is 6 ${\rm \mu}$m for most dimensions.  The parts are bonded together using a room-temperature hydroxide catalysis technique that locally dissolves the surfaces of the fused silica parts, fusing them together in a siloxane bond and liberating water in the process\cite{gwo}.  These bonds can withstand repeated cooling to $\sim$2 K when placed in the cryostat of Figure 2.  This technique facilitates precise component positioning with minimal added mass.

\begin{figure}[ht]
\centerline{\epsfxsize=2in\epsfbox{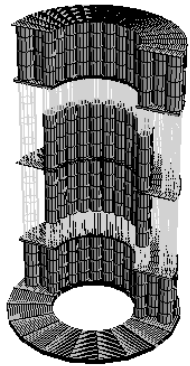}}   
\caption{Rendition of the ISLV source mass.
\label{islvsource}}
\end{figure}

The ISLV source mass is designed in a fashion similar to the pendulum and is being assembled at U. Washington.  All {\it m} = 1 Newtonian derivatives of the potential below {\it l} = 8 are eliminated.  The even {\it l}, {\it m} = 1 potentials are naturally zero because of source-mass top/bottom symmetry about the horizontal mid-plane.  Our design therefore has four degrees of freedom used to null the 331, 551, 771, and 220 potentials.  The 220 potential is eliminated to render the source mass insensitive to small tilts about a horizontal axis, which would generate a 221 moment.  The resulting design configuration is shown in Figure \ref{islvsource}.  The mass is assembled by stacking several hundred precision-machined solid stainless steel cylinders with a mass totaling about 1500 kg.

Several trim masses are located at the outer radius of the stack in order to remove the residual 221 and 331 field moments resulting from design deviations in the fabrication of the cylinders or their stacking.  If necessary, the 441 field moment can also be trimmed.  The measurement of these residual fields is explained below.

The segmentation of the source mass into these many small cylinders allows us to measure and correct for small mass differences resulting from machining tolerances and density variations.  The ends of the cylinders have shallow male/female mating interfaces.  As each cylinder must be placed to within 50 ${\rm \mu}$m of the design position, a 15 ${\rm \mu}$m lateral clearance in the interface allows for small corrections in the placement of each piece as the stack is assembled.  The precise placement of cylinders is monitored during assembly in real time with laser-coordinate-measuring equipment.  Hollow stainless steel tubes provide structural  spacers and support the central and top portions of the stack.

\subsection{Experimental Technique}

The torsional motion of the pendulum is tracked using electro-optical techniques.  The source mass, which rests upon an air bearing, is rotated through a series of static positions in azimuth.  At each position the motion of the pendulum through several torsional cycles is recorded and fit to a multi-parameter model.  An azimuthal \break {\it m} = 1 variation in torque on the pendulum would constitute an ISLV signal and be manifest as a corresponding variation in several of these parameters.  As discussed in the introduction, commonly measured indicators of a torque would be an {\it m} = 1 \break variation in pendulum equilibrium orientation (deflection method) or torsional oscillation frequency (frequency method).  Each of these detection methods has advantages and drawbacks.

When initially operating this experiment at room temperature, we will instead measure the {\it m} = 1 variation in the amplitude of the second harmonic of the pendulum motion as our signal (``second-harmonic method")\cite{uwa2}.  This choice is made for two reasons.  First, the amplitude of the second harmonic is insensitive to temperature variations of the fiber\cite{uwmoriond}.  Second, the second harmonic amplitude is insensitive as well to tilt of the fiber attachment point.  As mentioned earlier, these effects can be limitations when using the deflection or frequency methods.

The temperature and tilt insensitivities of the second-harmonic method come at the possible cost of a longer integration time relative to the frequency method if the measurements are dominated by additive white noise (e.g. shot noise) rather than thermal noise of the pendulum system.  This is partly because white noise on the second harmonic amplitude scales inversely as the number of periods to the ${\rm 1/2}$ and not to the ${\rm 3/2}$ power as for the frequency method.  Additionally, measurement of a signal torque using the second-harmonic method is intrinsically nearly an order of magnitude more sensitive to white-noise timing error than when the frequency method is used, even for a single oscillation period.  At room temperature, however, using the frequency method would require the fiber temperature to be held constant to roughly a ${\rm \mu}$K to achieve the measurement goal of this experiment, a severe challenge.  However, we plan to repeat the experiment at the Hanford site using the cryogenic apparatus described earlier.  When operating at $\sim$2 K, the reduced temperature coefficient of the torsion fiber and improved temperature regulation should enable us to reap the strong advantage of shortened integration time.

The room temperature pendulum is enclosed in a vacuum chamber evacuated to a pressure below ${\rm 10^{-6}}$ mbar where the damping of the torsion oscillations is dominated by the fiber and not by  gas in the chamber.  A magnetic damper suppresses low frequency (non-torsional) mechanical modes of the pendulum--fiber system.  A three-layer magnetic shield demonstrably renders magnetic effects negligible.  The instrument is suspended from a three-axis passive vibration isolation system with characteristic frequency below 1 Hz.

\begin{figure}[ht]
\centerline{\epsfxsize=4in\epsfbox{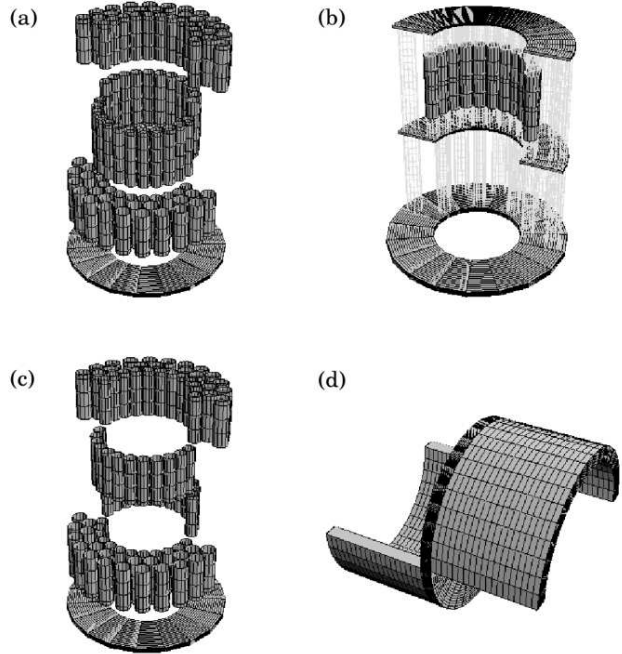}}   
\caption{(a) Rendition of the exaggerated 221 source mass.  (b) Rendition of the exaggerated 331 source mass.  The support structure, composed of hollow tubes and trays similar to those in the standard ISLV configuration, are shown.  (c) Rendition of the exaggerated 441 source mass.  (d) Rendition of the exaggerated 221 pendulum.
\label{s21}\label{s31}\label{s41}\label{pend21}}
\end{figure}

The residual 221 and 331 mass moments of the pendulum, which result from the non-zero dimensional tolerances associated with the fabrication process, are suppressed by four movable trim masses attached to the outer face of the thick disk shown in Figure \ref{islvpend}.  These critical mass moments are first measured by stacking the source mass in configurations that exaggerate the potentials coupling to these moments.  Figures \ref{s21}(a), \ref{s21}(b), and \ref{s21}(c) show the 221, 331, and 441 exaggerated configurations  respectively.  The 441 moment of the pendulum is not trimmed, but is measured to verify that it falls below the maximum acceptable value.

The lowest-order field moments of the source mass are also measured and if neccessary trimmed to conform to the design goals.  The pendulum used for this trimming procedure has an exaggerated 221 mass moment and is shown in Figure \ref{pend21}(d).  This pendulum is used directly to measure and trim the residual 221 field moment.  Moreover, because
\begin{equation}
V_{331} = \frac{\partial}{\partial z} V_{221}
\end{equation}
and
\begin{equation}
V_{441} = \frac{\partial^2}{\partial z^2} V_{221}
\end{equation}
the residual 331 and 441 moments can be measured with the 221 pendulum alone by measuring the 221 field at several elevations relative to the mid-plane of the source mass.  The linear and quadratic variation of the signal with elevation then yields the magnitude of the 331 and 441 fields.  Once these fields are measured, they can be trimmed away iteratively using the trim masses located at the back of the source mass stack.

\subsection{The Design Goals}

Uncompensated fabrication errors in the ISLV pendulum and source mass will result in spurious apparent values of the measured parameter ${\it \alpha}$ in Equation \ref{fullpot}.  To assess the resulting error in the measurement of ${\it \alpha}$ we first estimate the magnitudes of the fabrication errors based on measurements of the completed source-mass cylinders, the measured performance of our 3-D laser coordinate measuring equipment, and practical experience in manufacturing fused silica parts.  For the source mass, 50 ${\rm \mu}$m placement errors of the individual cylinders dominate the 6 ${\rm \mu}$m fabrication errors whereas for the pendulum it is the 6 ${\rm \mu}$m thickness tolerances of the fused silica components that dominate.  These placement and machining errors are not treated as RMS values because we do not assume that the individual errors will be entirely uncorrelated.  Instead, worst-case error estimates for mass and field moments were based on these placement and fabrication errors as maximum deviations from the three-dimensional design configuration.  Once the apparatus is fully assembled, and we are able to empirically measure the 221, 331, and 441 mass moments and field moments, the measured values should be less than these worst-case estimates.

To gauge the magnitude of the fundamental constraints posed by systematic error from residual Newtonian gravitational coupling between pendulum and source mass, Table \ref{errsig} gives the equivalent contribution to ${\it \alpha}$ (at a range ${\it \lambda}$ = 12 cm) for interactions resulting solely from the worst-case fabrication and placement errors.  We see that the limiting interactions are the 441 and 771 terms at a few times ${\rm 10^{-6}}$.  Aside from these limiting systematic errors, there is also the statistical measurement uncertainty.  For practical integration times at room temperature, we may be able to constrain ${\it \alpha}$ to less than ${\rm 10^{-4}}$.  When the experiment is repeated at $\sim$2 K, the thermal noise is expected to be significantly reduced, possibly allowing an upper limit on ${\it \alpha}$ below ${\rm 10^{-5}}$.  Cryogenic operation may effectively remove noise considerations and leave the measurement limited by the Newtonian gravitational interactions.  Such a limit is implied by Table \ref{errsig}, and would yield a constraint on ${\it \alpha}$ of about 5 x ${\rm 10^{-6}}$, limited by the worst-case 441 interaction.  It may be possible to trim the 441 moment of the source mass by as much as 90\% leaving the worst-case 771 interaction as the limiting systematic effect. 

\begin{table}[bt]
\tbl{Gravitational second harmonic signals arising from the combined fabrication errors of the pendulum and source mass.}
{\footnotesize
\begin{tabular}{@{}cc@{}}
\hline
{} &{} \\[-1.5ex]
Multipole Signal & Equivalent ${\it \alpha}$ (${\it \lambda}$ = 12 cm) \\[1ex]
\hline
{} &{} \\[-1.5ex]
221 & 8.22 ${\times 10^{-7}}$ \\[1ex]
331 & 1.32 ${\times 10^{-7}}$ \\[1ex]
441 & 4.60 ${\times 10^{-6}}$ \\[1ex]
551 & 3.58 ${\times 10^{-7}}$ \\[1ex]
771 & 1.31 ${\times 10^{-6}}$ \\[1ex]
991 & 3.23 ${\times 10^{-8}}$ \\[1ex]
\hline
\end{tabular}\label{errsig} }
\vspace*{13pt}
\end{table}

\section*{Acknowledgments}
We would like to thank and to acknowledge the support of the Pacific Northwest National Laboratory, our experimental site host; Los Alamos National Laboratory for fabrication of our {\it G} source masses; the National Institute of Standards and Technology for metrology of our {\it G} source masses; and the Boeing Company for the long-term loan of state-of-the-art laser coordinate measuring equipment.  This research is funded under National Science Foundation grants PHY-0108937 and PHY-9803765.

%
%
%
%

\end{document}